\documentclass[useAMS,usenatbib]{mn2e}

\usepackage{graphicx}
\usepackage{amsmath}
\usepackage{amssymb}
\usepackage{color}
\usepackage{gensymb}
\usepackage{subfigure}

\title[Transits Probabilities Around HVSs and RSs]{Transit Probabilities Around Hypervelocity and Runaway Stars}
\author[G. Fragione and I. Ginsburg]{G. Fragione$^{1}$\thanks{E-mail: giacomo.fragione@uniroma1.it} and I. Ginsburg$^{2}$\\
$^{1}$Department of Physics, Sapienza, University of Rome, P.le A. Moro 2, Roma, Italy\\
$^{2}$Astronomy Department, Harvard University, 60 Garden St., Cambridge, MA 02138, USA}

\begin{document}

\maketitle

\begin{abstract}
In the blooming field of exoplanetary science, NASA's \textit{Kepler Space Telescope} has revolutionized our understanding of exoplanets. \textit{Kepler}'s very precise and long-duration photometry is ideal for detecting planetary transits around Sun-like stars. The forthcoming \textit{Transiting Exoplanet Survey Satellite (TESS)} is expected to continue \textit{Kepler}'s legacy. Along with transits, the Doppler technique remains an invaluable tool for discovering planets. The next generation of spectrographs, such as \textit{G-CLEF}, promise precision radial velocity measurements. In this paper, we explore the possibility of detecting planets around hypervelocity and runaway stars, which should host a very compact system as consequence of their turbulent origin. We find that the probability of a multi-planetary transit is 
$10^{-3}\lesssim P\lesssim 10^{-1}$. We therefore need to observe $\sim 10-1000$ high-velocity stars to spot a transit. However, even if transits are rare around runaway and hypervelocity stars, the chances of detecting such planets using radial velocity surveys is high. We predict that the European \textit{Gaia} satellite, along with \textit{TESS} and the new-generation spectrographs \textit{G-CLEF} and \textit{ESPRESSO}, will spot planetary systems orbiting high-velocity stars.
\end{abstract}


\label{firstpage}

\section{Introduction}

Discoveries of exoplanets have proliferated in the past decade primarily due to observations with the Doppler technique and transits. Thanks to the high precision achieved with today's spectrographs, Doppler spectroscopy allows for the determination of a planet's minimum mass based upon the shift of stellar absorption lines. \citet{cum08} analysed eight year's worth of radial velocity measurements for nearly $600$ FGKM stars. The fundamental observational quantity is the stellar velocity amplitude induced by the planet \citep{cum04,cum08}. \citet{cum08} showed that $17$-$20$\% of stars have gas giant planets within $20$ AU. \citet{may11} reported the results from an eight year survey using the HARPS spectrograph. They conclude that greater than half of solar-type stars harbour a planet with a period of $\leq 100$ days. Furthermore, they find that $\sim 14$\% of solar-type stars host a planet with mass greater than $50$ M$_{\bigoplus}$. Doppler observations are of vital importance and continue to help in discovering new planets (e.g. \citet{dai16}). However, today transits dominate the search for exoplanets.

A transit is the passage of a smaller body in front of a larger body, such as when an exoplanet passes in front of its host star thus producing a drop in brightness \citep{win10}. Several surveys have been dedicated to transits detections, the most important and fruitful of which is NASA's \textit{Kepler Space Telescope}, which has revolutionized exoplanetary science \citep{bor10}. 
\textit{Kepler}'s original purpose was to determine the frequency and characteristics of planets and planetary systems in the habitable zone around FGKM stars. However, Kepler's very precise and long-duration photometry is ideal for detecting systems with multiple transiting planets \citep{lis11}. NASA's next major exoplanet mission scheduled for launch in 2017 is the \textit{Transiting Exoplanet Survey Satellite (TESS)}. \textit{TESS} is expected to monitor several hundred thousand Sun-like stars for transiting planets across nearly the entire sky using four wide-field cameras \citep{ric15}. \textit{TESS} aims to combine the strengths of wide-field surveys with the fine photometric precision and long intervals of \textit{Kepler}, but compared to Kepler will examine stars that are generally brighter by $3$ mag over a solid angle that is larger by a factor of $\sim 400$ \citep{sul15}. In this paper, we explore the possibility of detecting planets around high-velocity stars using both the Doppler technique and transits. 

High-velocity stars are most often Galactic halo stars with high peculiar motions, usually divided in two different categories, runaway stars (RSs) and hypervelocity stars (HVSs). RSs are historically defined as Galactic young halo stars with peculiar motions higher than $40$ km s$^{-1}$, which are thought to have travelled to the halo from their birthplace. RSs are produced in binary systems thanks to dynamical multi-body interactions or due to the velocity kick from a supernova explosion \citep{sil11}. HVSs, on the other hand, are stars escaping the Galaxy. \citet{hil88} was the first to predict the existence of HVSs, while \citet{brw05} discovered the first HVS in the outer halo. Hills' mechanism involves the tidal breakup of a binary passing close to a massive Black Hole (BH) \citep*{gil06,gil07,brw15,frl16}. Other mechanisms have, also, been proposed to explain the existence of HVSs, as the interaction of a massive binary black hole with a single star \citep{yut03}, or the interaction of star clusters and BHs \citep*{cap15,fra16,fck16}. Observations of high velocity and hypervelocity objects have usually been limited to high-mass, early-type, stars, due to observational bias \citep*{brw14}. However, recently observers have begun investigating low-mass HVSs candidates \citep*{lii12,pal14,fav15}. The European Space Agency (ESA) satellite \textit{Gaia} is expected to measure proper motions with an unprecedented precision, providing a larger and less biased sample ($\sim 100$ new HVSs in a catalogue of $\sim 10^9$ stars). Moreover, \textit{Gaia}'s sensitivity is good enough to search for multi-planet systems around massive stars and evolved stars and reveal their architecture and three-dimensional orbits \citep{cas08,wif15}. Furthermore, the detection of a planet around a HVS or runaway star will provide valuable information on the survivability of planets in extreme environments \citep*{gin12}. 

Thus, in this paper we look at the likelihood of finding transits around such high-velocity stars. 
In Section 2 we discuss our approach to calculating transits including a discussion on the code we used. In Section 3 we explore various possibilities and discuss our outcomes. In Section 4 we explain additional difficulties that are inherent in observing multi-planet transits. 
In Section 5 we discuss the Doppler technique. We conclude with a discussion and implications for future observations in Section 6.

\section{Method}

The geometry of transits can be computed planet-by-planet only when dealing with planets independently \citep{mur10}. When multi-planet systems are considered, the geometry becomes more complicated and has to be correctly understood in order to infer information on the architecture of such systems. Moreover, the ideal geometrical case does not provide a correct value for the transit probability because it fails to account for non-Keplerian orbits, duty cycles and signal-to-noise ratios \citep{bra16}.

The geometric probability of a transit of a single planet \citep{mur10,win10} is given by
\begin{equation}
p_T=\frac{R_*\pm R_p}{a}\frac{1+e\sin\omega}{1-e^2},
\end{equation}
and depends on the star and planet radius ($R_*$ and $R_p$ respectively), the planet's orbital eccentricity ($e$),  and the argument of periapsis ($\omega$). The "+" sign allows grazing eclipses, while the "-" sign excludes them. In the case that $R_p\ll R_*$, and marginalizing over $\omega$, the probability is simply
\begin{equation}
p_T\approx\frac{R_*}{a(1-e^2)}\approx 0.005 \left(\frac{R_*}{\mathrm{R}_{\odot}}\right)\left(\frac{1\ \mathrm{AU}}{a}\right)\left(\frac{1}{1-e^2}\right)
\label{eqn:probab}
\end{equation}
A natural question is: how many systems do we need to observe to spot a transit? Once the planetary orbital distance is determined, we have to observe \citep{win10}
\begin{equation}
N\gtrsim(\eta\ p_T)^{-1}
\label{eqn:numbp}
\end{equation}
stars, where $\eta$ is the fraction of stars that are thought to host such planets.

While the geometry of a single planet transit allows for an analytical calculation, there is no analytical solution for the transit probability of $M$ planets. Furthermore, observations have shown that multi-planet systems are likely common. \citet{lis14} and \citet{row14} determined that approximately 40\% of \textit{Kepler}'s candidate are in multi-planet systems. Some of those systems are very compact. For example, \citet{lsf11} presented Kepler-11, a single Sun-like star with six transiting planets. Five of those planets have orbital periods between $\sim 10$ and $\sim 47$ days. \citet{lis12} studied the planetary system Kepler-33. In this system there are five transiting planets with periods ranging from around $5$ to $\sim 41$ days. Precise Doppler measurements have also revealed very compact systems. As an example, GJ $876$ \citep{mrc01,riv10} is a red dwarf with four planets orbiting the star with semi-major axis between $\sim 0.02$ AU and $\sim 0.33$ AU. As pointed out by \citet{lis12}, the vast majority of \textit{Kepler}'s candidate multiple transiting systems are planets. The false positive rate being less than $\sim 1$\%. While in many instances it is safe to assume a single-planet system since the probability of detection is dominated by the innermost planet, clearly that is not appropriate in such compact systems and thus numerical techniques are required.

\subsection{Multi-planet systems}

\begin{figure}
\centering
\includegraphics[scale=0.65]{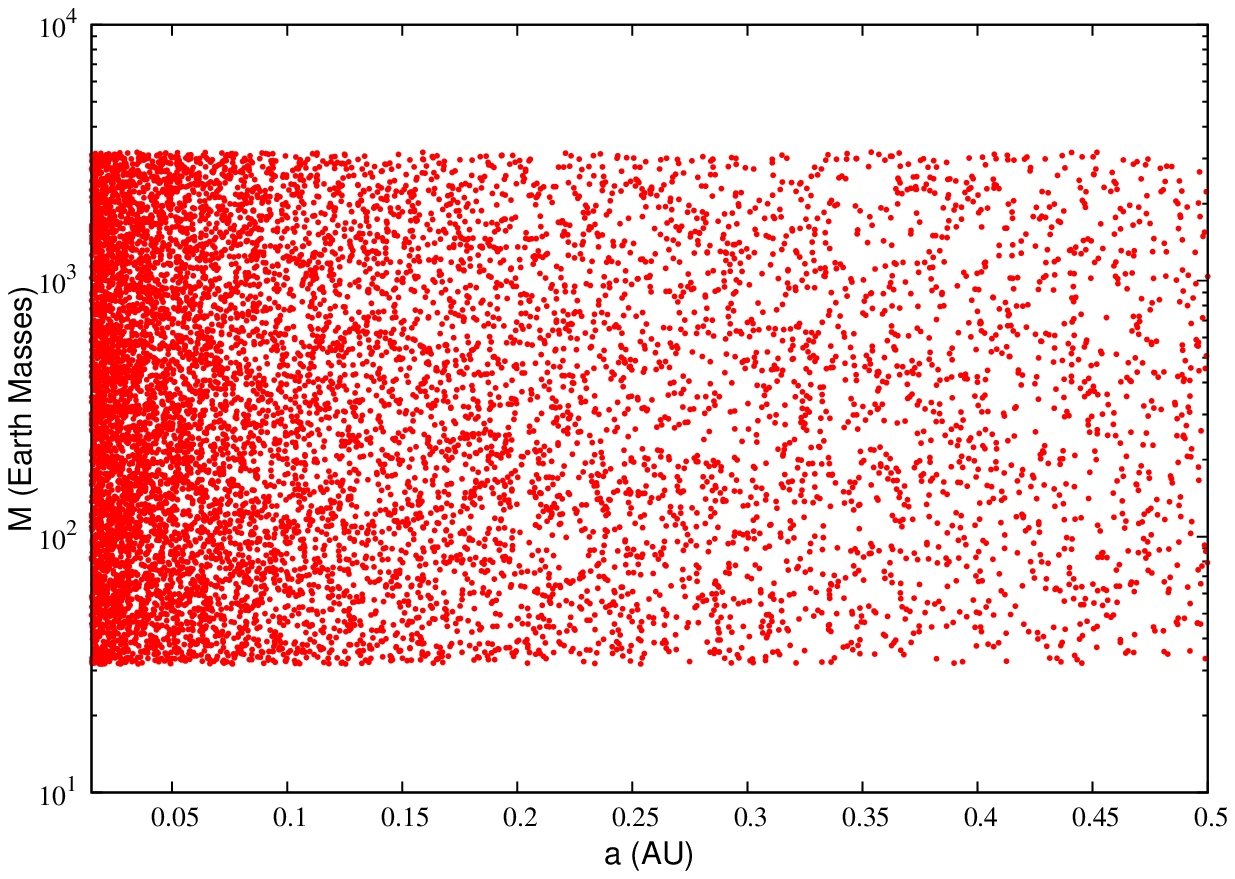}
\includegraphics[scale=0.65]{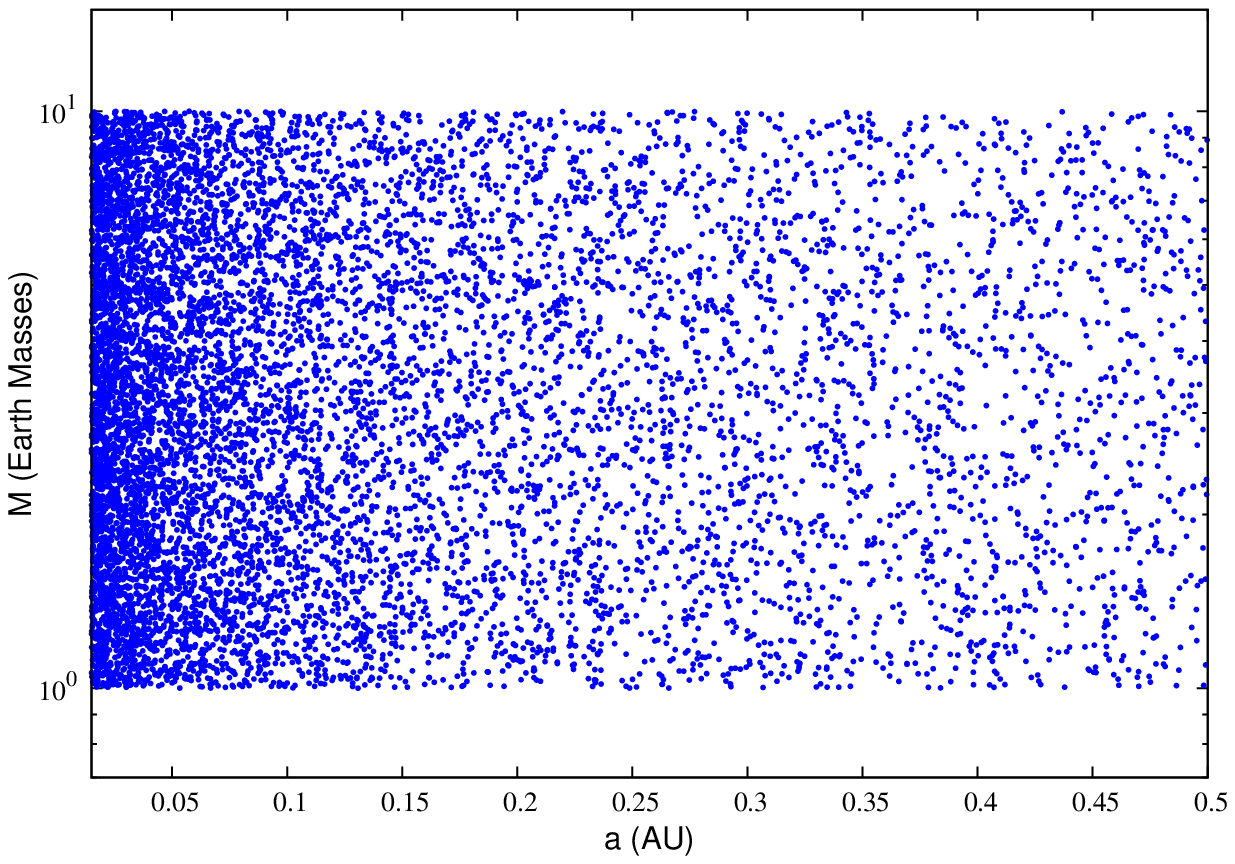}
\caption{Planetary parameters (masses and semi-major axis) for high mass stars ($M_* > 1$ M$_{\odot}$, top panel) and low mass stars ($M_* \le 1$ M$_{\odot}$, bottom panel).}
\label{fig:plandata}
\end{figure}

In order to compute the transit probabilities for multi-planet systems we use the publicly available code \textsc{CORBITS} \citep{bra16}. \textsc{CORBITS} is an algorithm that computes the combined geometric probability of a multi-transit in an exoplanetary system. The geometric probability of a transit is defined as the solid angle swept out by the planet's shadow on the celestial sphere. The transit region is defined as the union of all the shadowed star regions due to the planets' orbital motions. \textsc{CORBITS} is able to compute the surface area of any arbitrary intersection of these transit regions once supplied the star's radius $R_*$ and the orbital elements $a$, $e$, $i$, $\omega$, $\Omega$ (semi-major axis, eccentricity, inclination, argument of periapsis, and longitude of the ascending node respectively) of each planet. The code calculates the half-thickness $h_j=p_{T,j}$ (see Eq.\ref{eqn:probab}) for each planet, the geodesic curvature of each transit and, finally, thanks to the Gauss-Bonnet theorem, the joint probability of a multi-transit \citep{bra16}. The code assumes fixed Keplerian orbits. The dynamical interactions between the planets alter the pure Keplerian orbit and the timing of the transits (Transit Timing Variation, TTV) \citep{ago05,hol05}. However, such TTVs affect planetary phase, that are not relevant for $p_T$ \citep{win10}, and modifications to the orbits are negligible on observational timescales \citep{bra16}.

We choose the following set of initial conditions based upon \citet{jut08}:
\begin{itemize}
\item Eccentricity generated according to a Rayleigh distribution with mean $0 \le \sigma_e \le 0.4$
\begin{equation}
f(e)=\frac{e}{\sigma_e}\exp\left(\frac{e^2}{\sigma_e}\right);
\end{equation}
\item Inclination generated according to a Rayleigh distribution with mean $0^{\circ} \le \sigma_i \le 40^{\circ}$
\begin{equation}
f(i)=\frac{i}{\sigma_i}\exp\left(\frac{i^2}{\sigma_i}\right);
\end{equation}
\item Semi-major axis generated according to a uniform distribution in $\log a$ with $0.015 \le a/\mathrm{AU} \le 0.5$;
\item Period is computed using Kepler's Third Law
\begin{equation}
P=\frac{a(\mathrm{AU})^{1.5}}{\sqrt{M_*(\mathrm{M_{\odot}})}}\ \mathrm{yr};
\end{equation}
\item The argument of periapsis is generated according to a uniform distribution with $0^{\circ} \le \omega \le 360^{\circ}$;
\item The longitude of the ascending node is generated according to a uniform distribution with $0^{\circ} \le \Omega \le 360^{\circ}$.
\end{itemize}

Although observed exoplanetary systems have small relative inclinations, and large mutual $\sigma_i$ tend to make systems more dynamically unstable \citep{wif15}, we computed transits probabilities up to a mean inclination of $40^{\circ}$. We are primarily interested in systems that have undergone very strong gravitational interactions \citep*{mal11}, as in the Hills mechanism that generates HVSs \citep*{gin12}. For the same reason, the exoplanetary system needs to be compact in order that the star is able to retain its planets \citep{gin12}. In order that the exoplanetary system is stable for enough time to be observable, the spacing of exoplanetary orbits has to satisfy the criterion \citep{cha96,smi09,lis11}
\begin{equation}
\frac{a_{i+1}}{a_i}=\frac{2(3M_*)^{1/3}+\beta(m_{i+1}+m_i)^{1/3}}{2(3M_*)^{1/3}-\beta(m_{i+1}+m_i)^{1/3}}
\end{equation}
where $a_i$ and $m_i$ are the semi-major axis and mass of the i-th planet, respectively, and $M_*$ is the host star mass. The lifetime of a planetary system generally decreases with increasing system multiplicity and planets masses and orbital eccentricities. However, the stability of a planetary system increases with the initial spacing measured in terms of $\beta$. $\beta$ is a parameter that guarantees that the relative distance of two subsequent planets is big enough with respect to the mutual Hill radius
\begin{equation}
R_H=\left(\frac{m_{i+1}+m_i}{3M_*}\right)^{1/3}\frac{a_{i+1}+a_i}{2}.
\end{equation}
\citet{cha96} suggested that $\beta \gtrsim 10$ for $M > 3$ planets ensures the planetary stability over Gyrs. \citet{smi09} determined that a spacing of $\beta \gtrsim 8$ is sufficient for Myr stability in systems with $M > 5$ equal mass planets. In our calculation, we set $\beta=10$. The calculation of the Hill stability needs an assumption on planetary mass and radius. We generate planets masses according to a uniform distribution in $\log M_p$ \citep{jut08} in the range $0.1 \le M_p/\mathrm{M_J} \le 10$ for $M_*> 1$ M$_{\odot}$. Figure \ref{fig:plandata} (top panel) illustrates the planets masses and orbital semi-major axis used in this paper for $M_* > 1$ M$_{\odot}$ stars. For low-mass stars ($M_* \le 1$ M$_{\odot}$), we prefer the range $1 \le M_p/\mathrm{M_{\bigoplus}} \le 10$ for two reasons. First, the transit probabilities computed by \textsc{CORBITS} assumes that $R_*\gg R_p$, which is not true for $M_* \le 1$ M$_{\odot}$ if $0.1 \le M_p/\mathrm{M_J} \le 10$. Second, \textit{Kepler} observations suggest that there is a prevalence of Earth-sized planets around Sun-like stars \citep{pet13,bur15,gai16}. However, the choice of two mass intervals for exoplanets is just to consider realistic systems spaced accordingly the Hill's criterion, since the transit is independent on the planets radii, but $p_T\propto a^{-1}$ (see Eq.\ref{eqn:probab}). Figure \ref{fig:plandata} (bottom panel) illustrates the planets masses and orbital semi-major axis used in this paper for $M_* \le 1$ M$_{\odot}$ stars. The mass range is important when dealing with the possibility of spotting real transits (see Section 4). When we generate the semi-major axis, it is accepted assuming it satisfies the above stability criterion; otherwise it is sampled again from the distribution.

\section{Geometrical transits}

\begin{figure}
\centering
\includegraphics[scale=0.65]{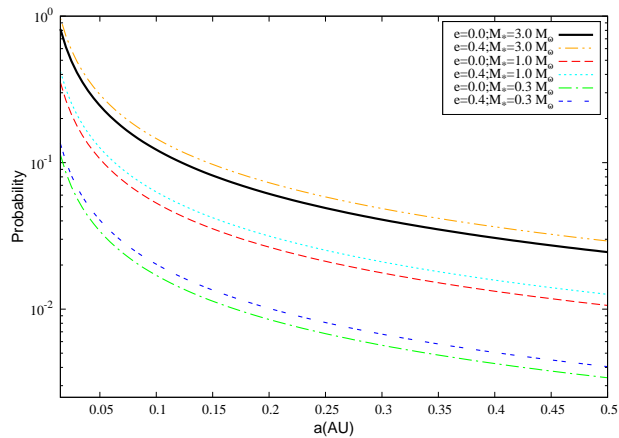}
\caption{Transit probabilities for an exoplanet orbiting a host star of mass $M_*=0.3$-$1.0$-$3.0$ M$_{\odot}$ as function of the orbital semi-major axis $a$.}
\label{fig:singplan}
\end{figure}

The geometry of transits can be computed analytically by means of Eq.\ref{eqn:probab} when dealing with stars that host a single exoplanet \citep{mur10,win10}. Figure \ref{fig:singplan} illustrates the transit probabilities for an exoplanet orbiting a star of mass $M_*=0.3$-$3.0$ M$_{\odot}$ as function of the orbital semi-major axis and eccentricity. The transit probability decreases as the orbital semi-major axis increases and is $10^{-3}\lesssim p_T\lesssim 1$. Moreover, the probabilities tend to be an increasing function of $e$ and of the stellar mass $M_*$ ($R_*\propto M_*^{\beta}$ with $\beta>1$, see below) since Eq.\ref{eqn:probab} is sensitive to high 
eccentricities and stellar radii. HVSs in the MMT survey are probably all main sequence $2.5-4$ M$_{\odot}$ B stars based on stellar rotation \citep{brw14}. Figure \ref{fig:singplan} shows that such massive stars yields large probabilities to planetary transits, which suggests that, if $\eta\sim 1$, about $50$ HVSs should be observed to spot a transit. The present sample of $\sim 20$ HVSs gives a probability of observing a transit $\sim 0.4$. Although massive stars lead to larger probabilities, a transit must take into account the relative flux decrement $\sim(R_p/R_*)^2$. 
For example, if $M_*=3.0$ M$_{\odot}$ it is possible that the star's luminosity would be overwhelming and thus a given telescope may not be sensitive enough to measure the drop in brightness (see Section 4).

\begin{figure}
\centering
\subfigure{\includegraphics[scale=0.65]{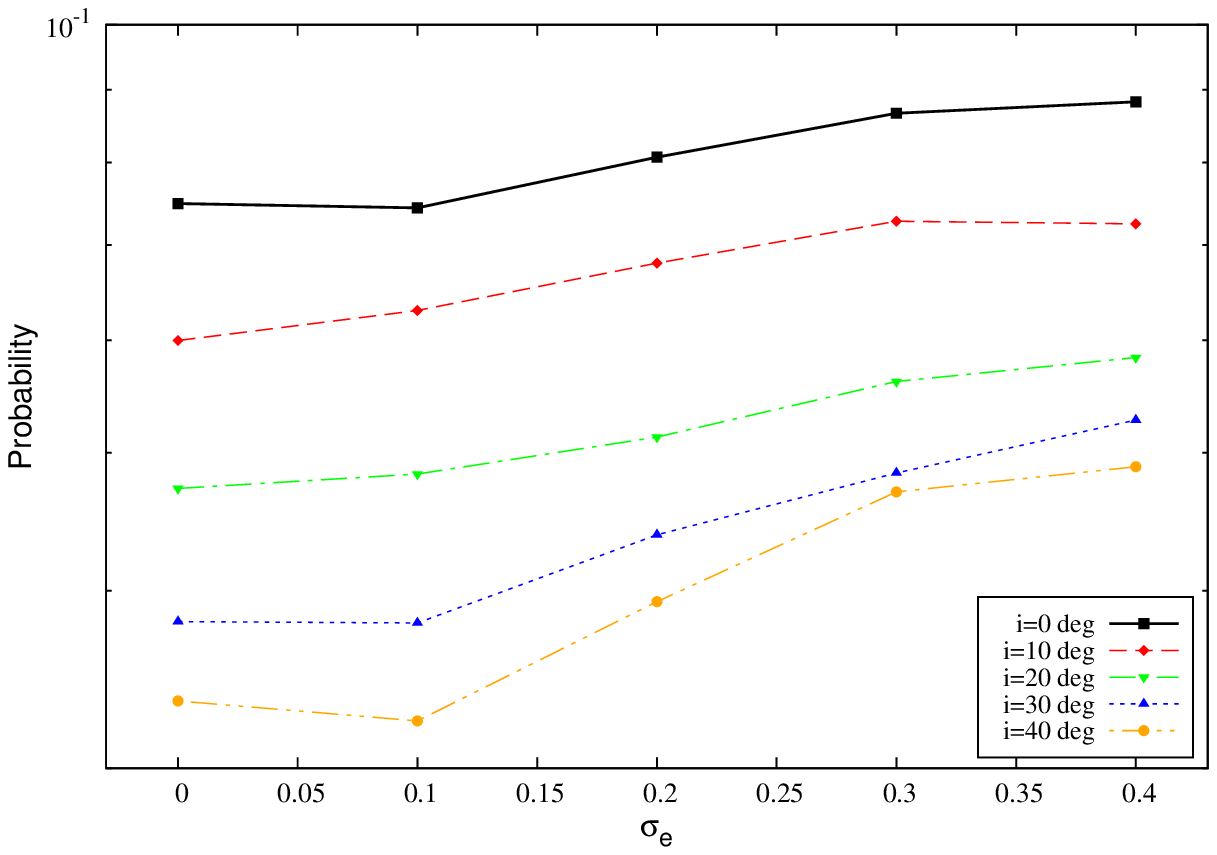}}
\subfigure{\includegraphics[scale=0.65]{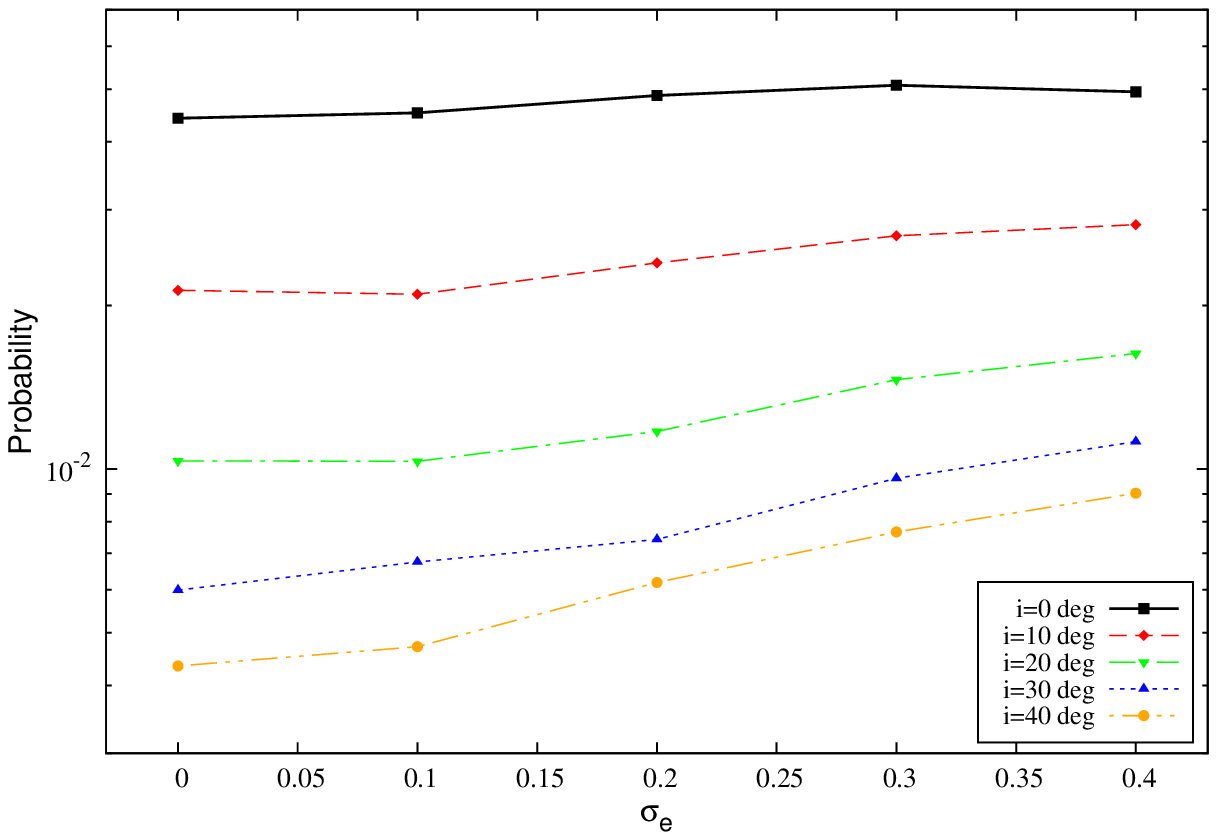}}
\subfigure{\includegraphics[scale=0.65]{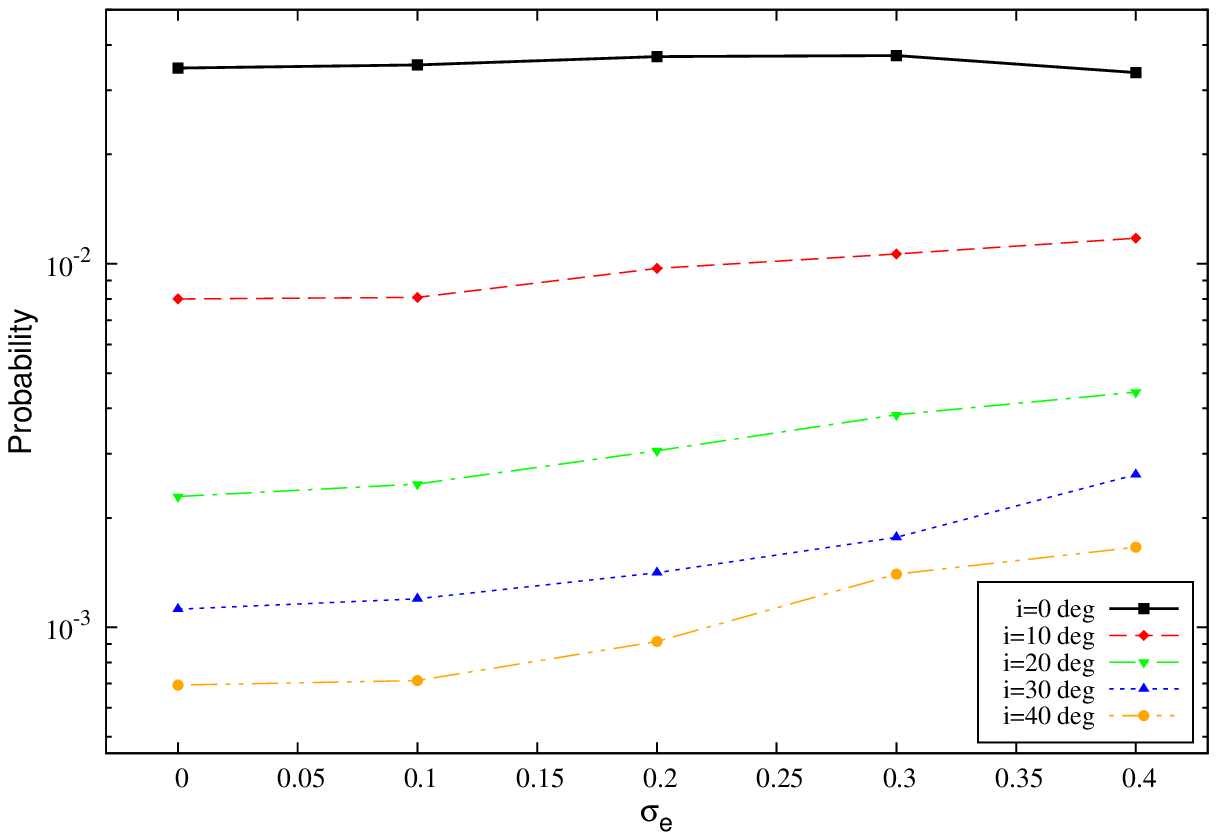}}
\caption{Transit probabilities for an exoplanetary system orbiting a host star of mass $M_*=3.0$ M$_{\odot}$ as function of the mean eccentricity $\sigma_e$ when the mean orbital inclination $0^{\circ}\le \sigma_i \le 40^{\circ}$. The number of planets varies from $2$ (top panel), $3$ (central panel) and $4$ (bottom panel) planets.}
\label{fig:3024pl}
\end{figure}

In the case of a multi-planetary system, given the initial conditions discussed in the previous section, we compute the transit probabilities with \textsc{CORBITS}. We consider different star masses $M_*$, number of planets $N_P$, mean planetary inclinations $\sigma_i$ and eccentricities $\sigma_e$. Each point in the figures (3, 4, and 5) is the average of $10^4$ probabilities. Figure \ref{fig:3024pl} shows the transit probabilities for an exoplanetary system orbiting a star of mass $M_*=3.0$ M$_{\odot}$ as a function of the mean orbital eccentricity and inclination. We vary the number of planets from $2$ (top panel), $3$ (central panel) and $4$ (bottom panel). Observed exoplanetary systems have small $\sigma_i$ since large mutual inclination tends to make systems 
dynamically unstable \citep{wif15}. However, we computed transits probabilities up to $\sigma_i=40^{\circ}$ since the planetary system can be highly perturbed after strong gravitational encounters \citep{mal11,gin12}. The same reason leads us to consider very compact 
systems ($0.015 \le a/\mathrm{AU} \le 0.5$) \citep{gin12}. The probability $P_T$ is a decreasing function of the mean inclination. Detecting all of the planets in multi-planet systems becomes unlikely even when mutual inclinations are $\lesssim 1^{\circ}$ \citep{bra16}. This illustrates one weak point in the transit method which is that from our perspective the planetary orbits need to 
line up with the star \citep{lis11,bra16}. However, probabilities tend to be an increasing function of $\sigma_e$ since Eq.\ref{eqn:probab} is sensitive to high eccentricities.

\begin{figure}
\centering
\includegraphics[scale=0.65]{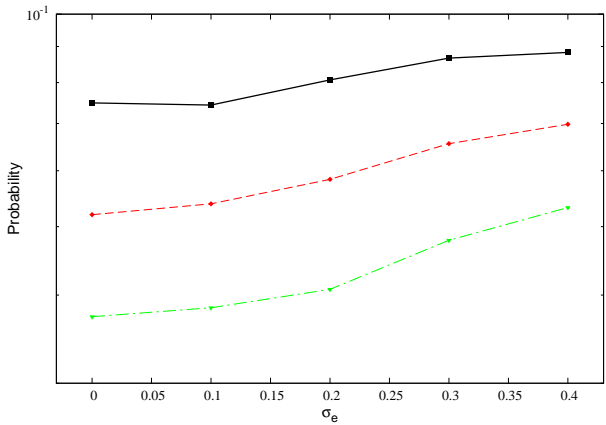}
\caption{Transit probabilities for an exoplanetary system orbiting a host star of mass $M_*=3.0$ M$_{\odot}$ as function of the mean eccentricity $\sigma_e$ when the mean orbital inclination $\sigma_i=0^{\circ}$ and the star hosts $2$ planets. The semi-major axis are sampled from a uniform distribution in $\log a$ (black line), an exponential distribution in $a$ (red line) and a uniform distribution in $a$ (green line).}
\label{fig:3024pldist}
\end{figure}

The joint probability of a multi-planetary transit is $10^{-3}\lesssim P_T\lesssim 10^{-1}$ and depends upon the semi-major axis range with $p_T\propto a^{-1}$. Equation \ref{eqn:numbp} is indicative of the number of systems we have to observe to spot a transit. If we assume that all the stars in our sample host such planetary systems, $\eta\sim 1$. Given the computed probabilities, we need to observe $\sim 10-1000$ stars to spot a multi-planetary transit. Here, we are considering only the geometrical probability which is larger then the real probability because of duty cycles and signal-to-noise (S/N) ratios (see Section 4). As discussed, HVSs in the MMT survey are probably
main sequence $2.5-4$ M$_{\odot}$ B stars \citep{brw14}. The present sample of $\sim 20$ HVSs gives a probability of observing a multi-planetary transit $\lesssim 0.2$ if $\eta\sim 1$. However, the \textit{Gaia} satellite is expected to find $\sim 100$ HVSs in a catalogue of $\sim 10^9$ stars \citep{ken14,brw15,deb15}. Assuming each such HVS hosts a compact planetary system of at least $2$ planets, Eq.\ref{eqn:numbp} predicts that at least one transit should be spotted.

Figure \ref{fig:3024pl} also illustrates the behavior of the transit probabilities as a function of the different number of planets, $N_P$. As $N_P$ increases, the probability of spotting a joint transit decreases. However, since the systems under consideration are constrained to be compact, and $p_T\propto a^{-1}$, the probability is $10^{-3}\lesssim P_T\lesssim 10^{-1}$, large enough to spot a system transiting one of the HVSs that are expected to be found with \textit{Gaia}. There are two additional effects to be considered. First, due to their origin few stars are expected to host a large number of planets \citep{mal11,gin12}. Even if the probability of spotting a transit of $4$ planets is $\sim 10^{-3}$, few high velocity stars are expected to have survived with several planets ($\eta \ll 1$) and hence the real probability of transits falls dramatically. Second, if $\sigma_i>0^{\circ}$, the transit probability can be smaller by an order of magnitude, or more for a higher number of planets. Systems with relative high inclinations are usually dynamically unstable, and observing a transit becomes quite difficult.

In our study, we have used a log-uniform distribution for the semi-major axis with $0.015 \le a/\mathrm{AU} \le 0.5$ \citep{jut08}. The details of the distribution function have important implications on the final joint probability. In order to investigate the effects of the semi-major axis distribution function, we computed transit probabilities sampling from different
distributions. Figure \ref{fig:3024pldist} shows the geometrical probabilities for $2$ planets orbiting a $3.0$ M$_{\odot}$ star ($\sigma_i=0^{\circ}$), when the semi-major axis are sampled from a log-uniform distribution (black line), an exponential distribution with mean $\lambda^{-1}=0.24$ AU (red line) and a uniform distribution (green line). Different semi-major axis distributions yield different geometrical probabilities, where the uniform distribution gives the lowest values. When sampling from that, large $a$, which yield to small $P_T$, have the same probability of small $a$, which yield to large $P_T$. On the contrary, the exponential distribution and the log-uniform distribution favor small semi-major axis, hence giving larger probabilities than the uniform distribution. However, the difference is just a factor of $\sim 2$. Similar results were obtained in the case that the $3.0$ M$_{\odot}$ star hosts $3$ or $4$ planets. Furthermore, simulations by \citet{gin12} show that the more compact a system, the more likely that system is to retain its planets after being disrupted. This suggests that a log-uniform or exponential distribution are well suited for our sampling.

\begin{figure}
\centering
\includegraphics[scale=0.65]{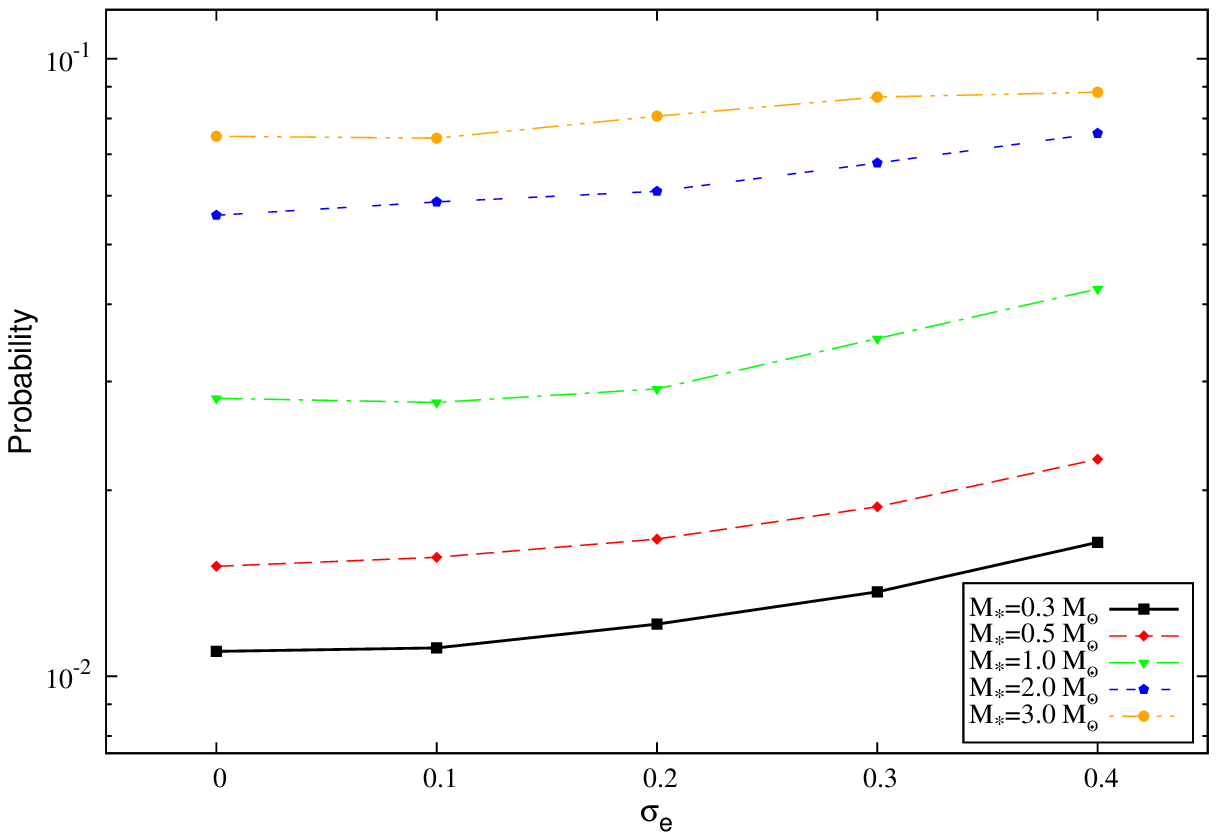}
\includegraphics[scale=0.65]{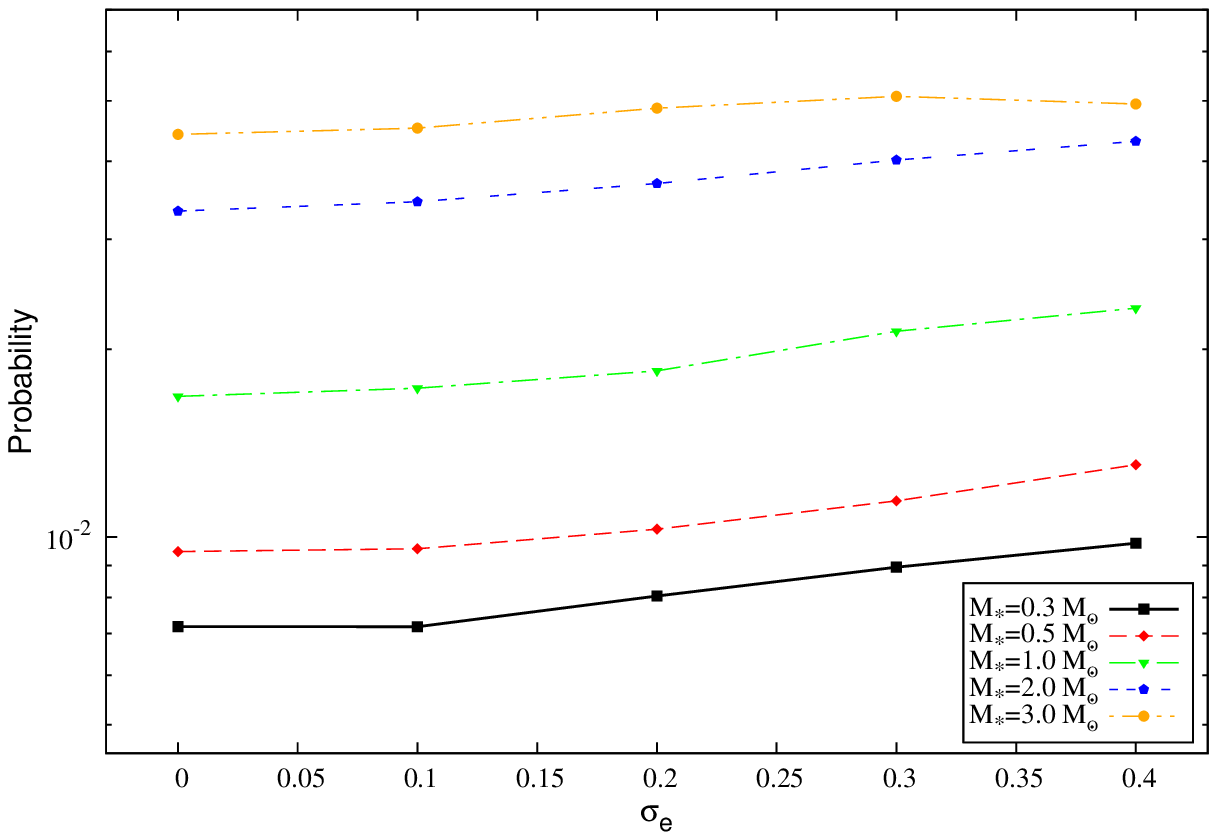}
\includegraphics[scale=0.65]{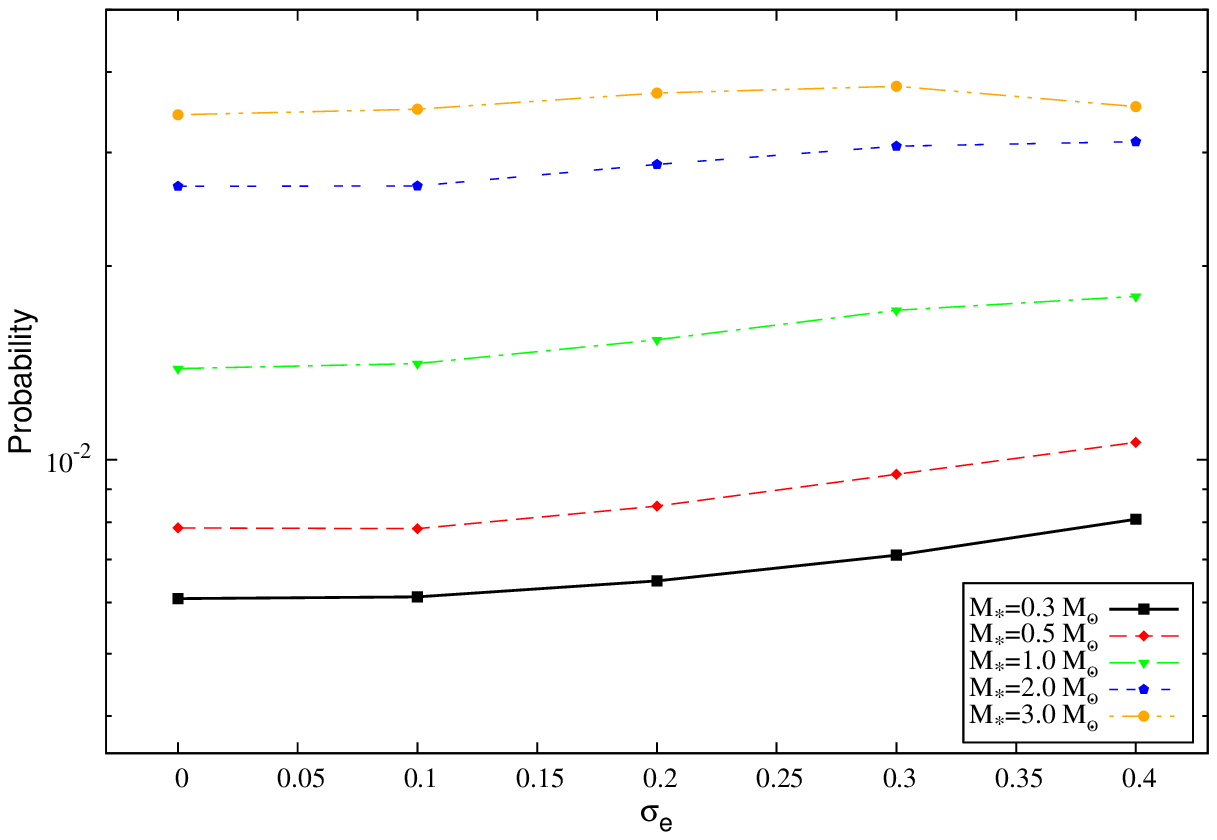}
\caption{Planets transit probabilities for an exoplanetary system orbiting a host star of mass $0.5 \le M_*/\mathrm{M}_{\odot} \le 3.0$ as function of the mean eccentricity ($\sigma_e$) when the mean orbital inclination $\sigma_i = 0^{\circ}$. We consider cases where the star hosts $2$ (top panel), $3$ (central panel), and $4$ (bottom panel) planets.}
\label{fig:masspl}
\end{figure}

Figure \ref{fig:masspl} shows the transit probabilities for an exoplanetary system orbiting a host star of mass $0.5\ \mathrm{M}_{\odot} \le M_*\le 3.0$ M$_{\odot}$ as function of the mean eccentricity ($\sigma_e$) when the mean orbital inclination $\sigma_i = 0^{\circ}$. We consider cases where the star hosts $2$ (top panel), $3$ (central panel), and $4$ (bottom panel) planets. The probability $P_T$ of a joint transit is an increasing function of the mass of the host star. Given the star's mass, we compute its radius from \citep{dem91}
\begin{equation}
R_*=
\begin{cases}
1.06\ M_*^{0.945}& \text{$ M_*< 1.66\ \mathrm{M}_{\odot}$},\\
1.33\ M_*^{0.555}& \text{$ M_*> 1.66\ \mathrm{M}_{\odot}$}.
\end{cases}
\end{equation}
Since $p_T\propto R_*$ and $R_*\propto M_*^{\beta}$, the geometrical probability of a transit increases with the star's mass. Even if massive stars lead to larger geometric probabilities, a transit must take into account the relative flux decrement $\sim(R_p/R_*)^2$, as discussed in the following section.

\section{Observing real transits}

Once we determine that a transit has a reasonable geometric probability, we have to address the question whether such a transit can be detected. The additional combined effect of duty cycles and S/N ratios must be taken into consideration \citep{bra16}.

The ratio $\Theta$ between the observing baseline of a telescope and the planets' orbital periods plays a fundamental role in detecting a transit. The transit duration of a single planet \citep{win10}
\begin{equation}
T=\left(\frac{R_* P}{\pi a}\sqrt{1-b^2}\right)\frac{\sqrt{1-e^2}}{1+e\sin\omega}\simeq 6\ \frac{P}{1\ \mathrm{day}}\frac{R_*}{a}\sqrt{1-e^2}\ \mathrm{hr}
\end{equation}
depends on the impact parameter $b$, which is the minimum projected distance between the star and planet, expressed in units of the stellar radius ($R_*$), the exoplanet orbital eccentricity ($e$), the argument of periapsis ($\omega$), and the period ($P$). If such a period is longer than the observing baseline, the probability of observing a single planet transit will become $p_{T,\theta}\approx p_T \Theta$. When dealing with multiple planets, the effect depends on the specific positions of the planets and the times of their observations. In this case it could be that $\Theta\ll 1$ and, even if the multi-transit has a reasonable geometric probability. Thus, we may be unable to observe such a transit.

To observe a transit, a photometric precision better than $\sim 1$\% is needed \citep{bea08}. De facto this requirement implies that transits must be relatively nearby. \textit{Kepler} has found planets orbiting stars out to a distance $\sim 2$-$3$ kpc. However, to quantitatively estimate if a transit can be observed, it is necessary to compute the S/N ratio. Even when a transit occurs, and $p_{T,\theta}$ is high, it might not be detectable because of a low S/N ratio. The signal, i.e. the drop in brightness caused by the passage of planets, is proportional to the decrease in intensity \citep{win10}
\begin{equation}
f(t)=\frac{F(t)}{F_*}=1+k^2\frac{I_p(t)}{I_*}-k^2\alpha_T
\end{equation}
which depends on the disk-averaged intensities $I_p(t)$ and $I_*$, the conversion factor $k\propto R_p/R_*$, which is proportional to the ratio of the planet and stellar radii ($F_p/F_*=k^2 I_p(t)/I_*$), and the dimensionless function $\alpha_T$, which depends on the overlap area between the stellar and planetary disks. The maximum loss of light is $f_{max}\simeq k^2\propto R_p^2/R_*^2$. However, because of limb darkening, $f_{max}$ is larger than $k^2$ when the planet is near the center of the star, and smaller than $k^2$ when the planet is near the limb. In any case, the computation of the S/N ratio for a transiting exoplanet is quite complicated, since one has to 
compute whether the transit would be detectable with specific transit-search pipelines \citep{fre13}. We assume a simple formula for converting planets masses into radii \citep{lis11}
\begin{equation}
R_p=M_p(\mathrm{M_{\bigoplus}})^{0.485}\ R_{\bigoplus}.
\end{equation}
If we consider a Jupiter-sized planet and a HVS of typical mass $3$ M$_{\odot}$, the flux decrement is proportional to $k\propto (R_J/R_*)^2\sim 0.1\%$. If the stellar mass is $\lesssim 1$ M$_{\odot}$, $k\sim 1\%$, while if the planet is Earth-sized, $k\lesssim 0.01\%$. Assuming that statistical correlation among data points from different transits are much weaker than among data points
during the same transit, \citet{pon06} obtained
\begin{equation}
\mathrm{S/N}=\frac{\Delta*n}{\sqrt{\sum_{k=1}^{N_{tr}}\left[n_k^2\left(\frac{\sigma_w^2}{n_k}+\sigma_r^2\right)\right]}},
\end{equation}
where $\Delta$ is the transit depth in magnitudes, $N_{tr}$ is the number of transits, $n$ is the number of data points observed during all transits, $n_k$ is the number of data points observed during the k-th transit, and $\sigma_r$ and $\sigma_w$ are the red and white noise in magnitudes, respectively. Whereas the white noise is uncorrelated from data point to data point, and typical sources are the photon and sky background noise, red noise is correlated from data point to data point and sources of red noise may be weather, seeing changes or intrinsic astrophysical changes in target brightness \citep*{pon06,von09}. The S/N ratio is thus a function of both the transit
survey strategy and of astrophysical parameters. Given the above quantities, if S/N exceeds a certain threshold value, a transiting planet is defined to be detectable in the data \citep{von09}. \citet{fre13} used $\chi=7.1$ as threshold for the \textit{Kepler} observations, while \citet{sul15} proposed a slightly larger threshold ($\chi=7.3$) for \textit{TESS} data. Even if detectable, the transit may not be observed. The window function determines the probability that a requisite number of transits required for detection occurs in the observational data, i.e. S/N transit exceeds this threshold, as a function of planetary orbital period \citep{bur06,von09,bur14,bur15}.
\citet{bea08} describe a general method to statistically calculate the number of transiting planets a given survey could detect. Calculating such transits is extremely complicated and often requires Monte-Carlo simulation. Such a treatment is beyond the scope of this paper. However, \citet{bur15} derived a simpler analytical expression of the form
\begin{equation}
\mathrm{S/N}=\sqrt{N_{tr}}\ \frac{\Delta}{\sigma},
\end{equation}
where $N_{tr}=M\times f_{duty}=(T_{obs}/P)\times f_{duty}$ is the expected number of transit events, $\sigma$ is the detection noise, $T_{obs}$ is the time baseline of observational coverage for a target and $f_{duty}$ is the observing duty cycle, defined as the fraction of $T_{obs}$ with valid observations. The probability of detection is defined as \citep{fre13,bur15}
\begin{equation}
P_{det}=\frac{1}{2}+\frac{1}{2}\ \mathrm{erf}\left(\frac{\mathrm{S/N}-7.1}{\sqrt{2}}\right),
\end{equation}
with a detection threshold of $7.1$. Given that, $50$\% of transits with $\mathrm{S/N}=7.1$ are detected, while the detection probabilities are $2.3$\%, $15.9$\%, $84.1$\%, $97.7$\%, and $99.9$\% for S/Ns of $5.1$, $6.1$, $8.1$, $9.1$ and $10.1$, respectively \citep{fre13}. Assume for simplicity $\Delta\sim R^2_p/R^2_*$ and $N_{tr}\sim 1$, if $\sigma\sim 1$\% the detection probabilities 
$2.3$\%, $15.9$\%, $84.1$\%, $97.7$\%, and $99.9$\% will correspond to $R_p/R_*\sim 0.23$, $0.25$, $0.28$, $0.30$ and $0.32$. Hence a Jupiter-sized planet transiting a $1$ M$_{\odot}$ runaway star has $\gtrsim 90$\% probability to be observed. On the other hand, if $\sigma\sim 10$\%, a $97.7$\% probability of detecting the transit implies that the host star and the planet have similar radii. 
When dealing with massive stars, such as the HVSs found by \citet{brw14}, $\sigma\gtrsim 0.01$ since HVSs are brighter then solar mass stars and more distant ($50\ \mathrm{kpc}\lesssim d \lesssim 120\ \mathrm{kpc}$). However, the S/N ratio can be significantly higher when dealing with the compact systems studied in the present work, since $\Delta$ will be the combination of the different $R^2_{p,i}/R^2_*$. Moreover, the \textit{Gaia} mission will be able to spot $\sim 100$ HVSs within few kpc from the Sun and without restrictions in star mass. Hypervelocity and runaway candidates (\citet{pal14}, \citet{fav15}, \citet{lii12}) may host planetary transits that can be 
observed with a $>50$\% probability according to the previous discussion. Finally, as discussed, in order to determine the final probability of observing a transit, the window function must be taken into account \citep{von09,bur15}. \citet{bur14} studied different window functions and found that planets with period $\lesssim 10$ days (as in the case of the planetary systems studied here) have a 
detection probability $\gtrsim 50$\%.

\section{Radial Velocities}

Along with transits, the Doppler technique can help find planets around high-velocity stars. Radial Velocity (RV) surveys search for the variance in time of the Doppler shift of absorption lines in stellar spectra resulting from orbital motion of the stars due to an exoplanet \citep{cla14}. In order to detect a planet, the measurement uncertainties have to be low enough to distinguish the periodic variation of the signal. The typical uncertainty of the measurement of a stellar absorption line \citep{bea15}
\begin{equation}
\sigma_{RV}\propto \frac{\Gamma^{3/2}}{W\ I_0^{1/2}}
\end{equation}
depends on the FWHM of the absorption line $\Gamma$, its equivalent width $W$ and the continuum intensity $I_0$. We are limited to luminous and nearby stars that have strong absorption lines and are bright enough to provide high S/N spectra \citep{ste13}.

The fundamental observational quantity in Doppler surveys is the stellar velocity amplitude induced by the planet \citep{cum08} 
\begin{eqnarray}
K&=&\frac{28.4\ \mathrm{m/s}}{\sqrt{1-e^2}}\frac{M_p \sin i}{\mathrm{M_J}}\left(\frac{P}{1\ \mathrm{yr}}\right)^{-1/3}\left(\frac{M_*}{\mathrm{M}_{\odot}}\right)^{-2/3}=\nonumber\\
&=&\frac{28.4\ \mathrm{m/s}}{\sqrt{1-e^2}}\frac{M_p \sin i}{\mathrm{M_J}}\left(\frac{a}{1\ \mathrm{AU}}\right)^{-1/2}\left(\frac{M_*}{\mathrm{M}_{\odot}}\right)^{-1/2}
\label{eqn:krad}
\end{eqnarray}
where $P$ is the orbital period, $e$ is the eccentricity, $M_*$ is the mass of the star, $M_p$ is the mass of the planet and $i$ is the 
orbital inclination. The S/N at which a RV survey can detect planets with a given period depends on the above stellar velocity amplitude,
the magnitude of the measurement uncertainties $\sigma$, the total number of observations $N_{RV}$ and the duration of the survey 
$T_{RV}$ \citep{cla14}
\begin{eqnarray}
\mathrm{S/N}&\approx& \left(\frac{N_{RV}}{2}\right)^{1/2}\left(\frac{K}{\sigma}\right)\nonumber\\
&\times&\left\{1-\frac{1}{\pi^2}\left(\frac{P}{T_{RV}}\right)^2\sin^2\left(\frac{\pi T_{RV}}{P}\right)\right\}^{1/2},
\label{eqn:snrad}
\end{eqnarray}
which, in the case $P\lesssim T_{RV}$, is well approximated by
\begin{equation}
\mathrm{S/N}\approx \left(\frac{N_{RV}}{2}\right)^{1/2}\left(\frac{K}{\sigma}\right).
\label{eqn:snradpt}
\end{equation}
Note that the above equation is nearly independent of the orbital period $P$. 
Given the above quantities, if S/N exceeds a certain threshold value $\chi_{RV}$, a RV planet is detectable in the data. The typical threshold depends on the duration of the RV survey. \citet{cum04} showed that, in the case $P< T_{RV}$, the threshold for 
$50$ per cent detection probability is 
\begin{equation}
\chi_{RV,PT}^{50}=\frac{1}{\sqrt{N_{RV}}}\left[4\ln\left(\frac{M}{F}\right)\right]^{1/2},
\end{equation}
and for $99$ per cent detection probability
\begin{equation}
\chi_{RV,PT}^{99}=1.7 \frac{1}{\sqrt{N_{RV}}}\left[4\ln\left(\frac{M}{F}\right)\right]^{1/2}.
\end{equation}
On the other hand, in the case $P> T_{RV}$, the threshold for $50$ per cent detection probability is
\begin{equation}
\chi_{RV,TP}^{50}=\frac{1}{\sqrt{N_{RV}}}\left[4\ln\left(\frac{M}{F}\right)\right]^{1/2}\left(\frac{2P}{\pi T_{RV}}\right),
\end{equation}
and for $99$ per cent detection probability
\begin{equation}
\chi_{RV,TP}^{99}=\frac{1}{\sqrt{N_{RV}}}\left[4\ln\left(\frac{M}{F}\right)\right]^{1/2}\left(\frac{2P}{\pi T_{RV}}\right)^2.
\end{equation}
In the previous equations, $M$ is the number of orbital frequencies ($2\pi/P_i$) that are used to describe the RV signal and $F$ is the false alarm probability. $M/F$ quantifies the significance of the Doppler signal due to an exoplanet based on how often a RV signal as large as the observed one would arise purely due to noise alone \citep{cum04,cum08}. Typical values are $M/F\sim 10^6$ \citep{cum04}. In the case of eccentric orbits, thresholds are more complicated \citep{bau15}. However, \citet{cum04} reported that the eccentricity has an important effect when $e \gtrsim 0.6$, where the threshold can be one or two order of magnitude larger.

\begin{figure}
\centering
\includegraphics[scale=0.65]{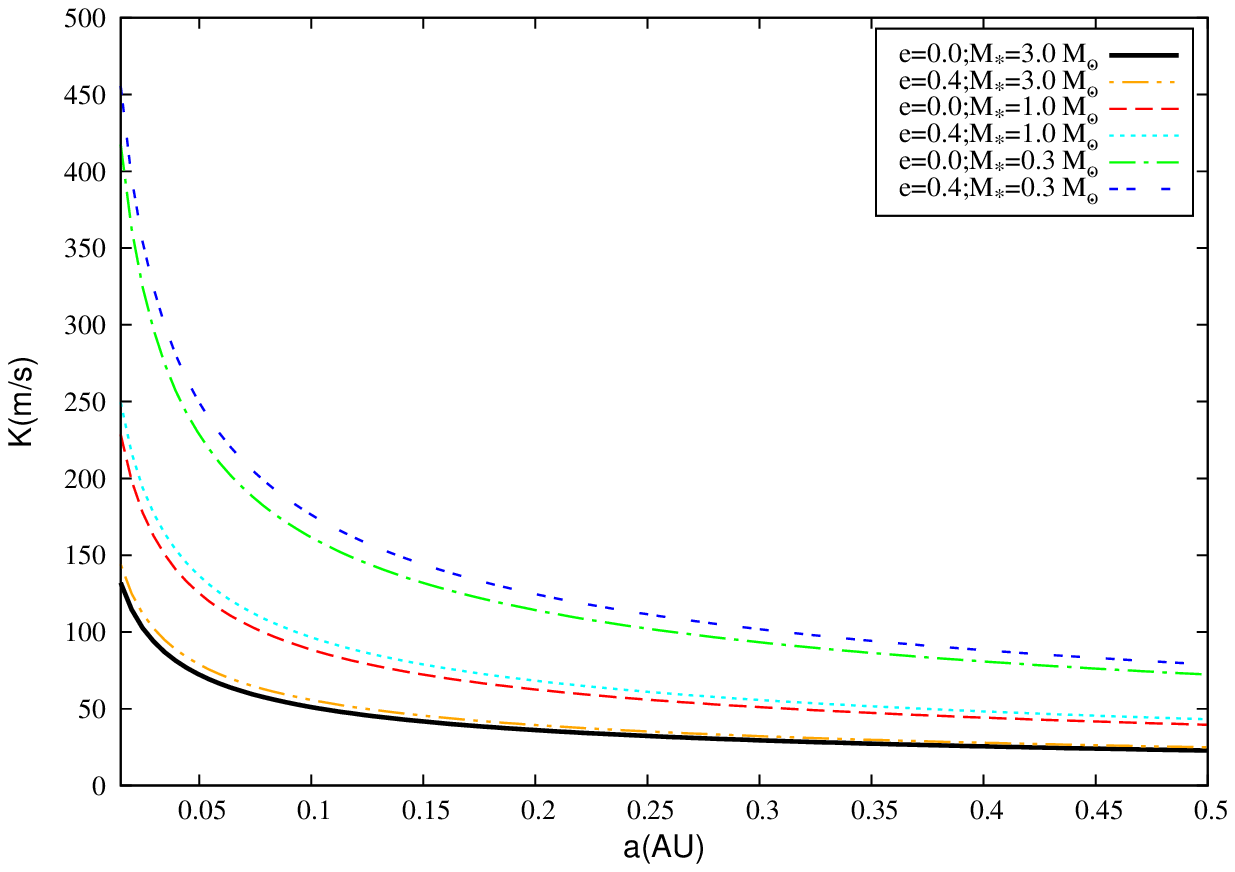}
\includegraphics[scale=0.65]{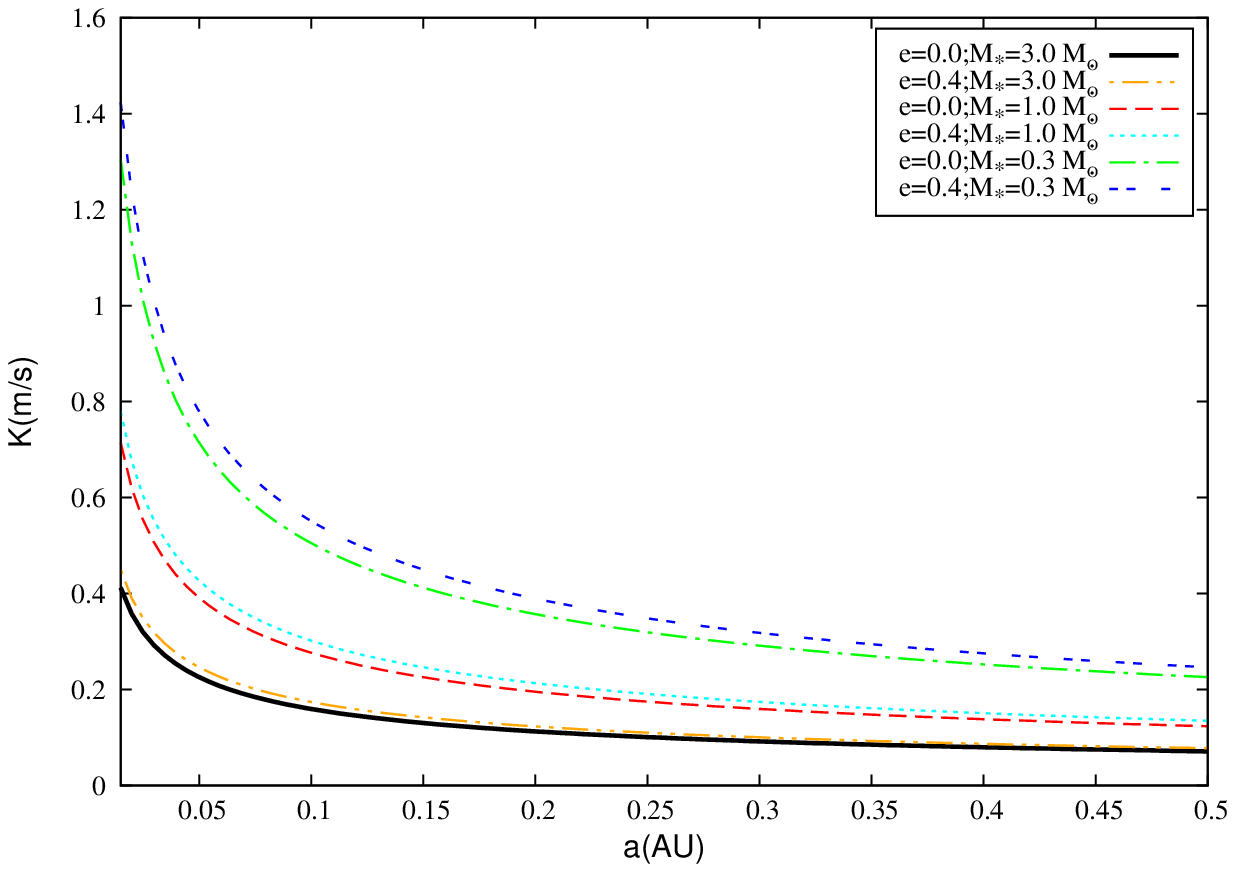}
\caption{Stellar velocity amplitude induced by a planet of minimum mass $1\ \mathrm{M}_\mathrm{J}$ (top panel) and $1\ \mathrm{M}_{\bigoplus}$ (bottom panel) orbiting a host star of mass $M_*=0.3$-$1.0$-$3.0$ M$_{\odot}$ as function of the orbital semi-major axis $a$.}
\label{fig:radv}
\end{figure}

Figure \ref{fig:radv} illustrates the stellar velocity amplitude induced by a planet of minimum mass $M_p \sin i=1\ \mathrm{M}_\mathrm{J}$ (top panel) and $M_p \sin i=1\ \mathrm{M}_{\bigoplus}$ (bottom panel) orbiting a host star of mass $M_*=0.3$--$3.0$ M$_{\odot}$ as function of the orbital semi-major axis $a$. Figure \ref{fig:radv} shows that the stellar velocity amplitude induced by the planet is $\sim 10$-$10^2$ m s$^{-1}$ for Jupiter-sized planets and of the order of unity for Earth-sized planets. The stellar velocity is also larger for smaller host stars and tight orbits, as consequence of Eq.\ref{eqn:krad}. Table \ref{tab:thres} illustrates the typical values of the thresholds $\chi_{RV}$ as a function of the number of observations, $N_{RV}$. In the case $P> T_{RV}$, such values depend also on the ratio of the orbital period and of the survey duration.

\begin{table}
\caption{Thresholds for radial velocity surveys}
\centering
\begin{tabular}{|c|c|c|c|c|}
\hline
$N_{RV}$ & $\chi_{RV,PT}^{50}$ & $\chi_{RV,PT}^{99}$ & $\chi_{RV,TP}^{50}$ & $\chi_{RV,TP}^{99}$\\
\hline
10  & 2.34 & 3.98 & 1.50$(P/T_{RV})$ & 0.96$(P/T_{RV})^2$ \\
\hline
50  & 1.05 & 1.79 & 0.67$(P/T_{RV})$ & 0.43$(P/T_{RV})^2$ \\
\hline
100 & 0.74 & 1.26 & 0.47$(P/T_{RV})$ & 0.30$(P/T_{RV})^2$ \\
\hline
\end{tabular}
\label{tab:thres}
\end{table}

Equations \ref{eqn:snrad} and \ref{eqn:snradpt} give the typical S/N ratio for a Doppler signal due to an exoplanet. As discussed, RV surveys usually target luminous and nearby FGKM stars \citep{ste13}. \citet{may11} reported the results from an eight year survey using the HARPS spectrograph with typical radial velocity accuracy $\sigma \sim 1$ m s$^{-1}$. New-generation spectrographs, such as \textit{G-CLEF} and \textit{ESPRESSO}, are expected to have $\sigma \lesssim 50$ cm s$^{-1}$. Using such fiducial values and results from Fig.\ref{fig:radv}, for a Jupiter-sized planet orbiting an FGKM star, S/N$\gtrsim 10^2$ well above the typical thresholds. On the other hand, Earth-sized planets have amplitudes, $K$, of the order of typical radial velocity accuracies and require long survey durations $T_{RV}$ and large $N_{RV}$, in particular if they orbit massive stars. \citet{cum04} showed that very eccentric planets ($e \gtrsim 0.6$) have thresholds that can be one or two order of magnitude larger. In this case, the RV would require very large $N_{RV}$, in particular for Earth-sized planets. As discussed, hypervelocity and runaway candidates (\citet{brw14}, \citet{fav15}, \citet{lii12}) may host compact planetary systems \citep{gin12}, that can be observed with a $>50$\% probability according to the previous discussion. In 
particular, RV surveys can help to spot the Doppler signal due to planets more massive then Earth that orbit low-mass FGKM stars.

\section{Conclusions}
\label{sec:con}

In this paper we have computed the likelihoods of finding exoplanets around high-velocity stars. We considered different stellar masses $M_*$, number of planets $N_P$, mean planetary inclinations $\sigma_i$ and eccentricities $\sigma_e$. We found that the geometrical probability of detecting a transit has generally an increasing trend with $\sigma_e$ as consequence of Eq.\ref{eqn:probab}, and decreases with $\sigma_i$. This indicates that the transit method may detect all of the planets in multi-planet systems if the planetary orbits are nearly lined up with the star \citep{lis11,bra16}. On the other hand, having a larger number of planets around less massive stars reduces
the probability of spotting a transit.

We considered a semi-major axis in the range $0.015 \le a/\mathrm{AU} \le 0.5$ since the planetary system must be compact to be retained after strong gravitational encounters \citep{gin12}. The joint probability of a multi-planetary transit is $10^{-3}\lesssim P\lesssim 10^{-1}$ and depends upon the semi-major axis range with $p_T\propto a^{-1}$. If we assume $\eta\sim 1$, Eq.\ref{eqn:numbp} predicts that we need to observe $\sim 10-1000$ stars to spot a transit. \textit{TESS} is expected to spot transiting planets across nearly the entire sky by monitoring several hundred thousand Sun-like stars \citep{sul15}. In particular, \textit{TESS} should be able to find transits around hypervelocity and runaway stars. For HVSs, the \textit{Gaia} satellite is expected to find $\sim 100$ such stars \citep{ken14,brw15,deb15}. If we assume that each HVS hosts a compact planetary system, Eq.\ref{eqn:numbp} predicts that at least one transit could be spotted.

Even if computations lead to larger geometric probabilities, a transit must take into account the relative flux decrement $(R_p/R_*)^2$ \citep{win10}. If we consider a Jupiter-sized planet and a HVS of typical mass $3$ M$_{\odot}$, the flux decrement is proportional to $k\propto (R_J/R_*)^2\sim 0.1\%$. If the star's mass is $\lesssim 1$ M$_{\odot}$, $k\sim 1\%$. While the geometrical 
probability favors heavier stars, the relative flux decrement indicates that the S/N ratio is larger in the case of small stars. Assuming $\sigma\sim 1$\%, a Jupiter-sized planet transiting a $1$ M$_{\odot}$ star have $\gtrsim 90$\% of probability to be observed. \textit{Gaia} will spot $\sim 100$ HVSs within few kpc from the Sun. Transits are likely to be observed around such HVSs with 
$\gtrsim 90$\% probability if they host planets.

Along with transits, the Doppler technique is an important tool for finding planets around high-velocity stars. Equation \ref{eqn:krad} indicates that the stellar velocity amplitude induced by the planet is $K\sim 10$-$10^2$ m s$^{-1}$, with larger values for FGKM stars. The detection threshold depends on the number and duration of the observations and on the noise sources \citep{cum08}. Generally the detection threshold is $\gtrsim 1$ m s$^{-1}$, and depends on the duration of the RV survey and on the planet's orbital period and eccentricity \citep{cum04}. 
The typical radial velocity accuracy ($\sigma \sim 1$ m s$^{-1}$) of modern spectrograph yield $> 50$\% probability of measuring the Doppler shift caused by a hot-Jupiter, whereas Earth-sized planets require long survey durations and large number of observations. In general, 
compact planetary systems with large planetary minimum masses around low-mass stars can generate a high Doppler signal, that can be measured with $>50$\% probability by the next-generation spectrographs.

As discussed, transits are particularly suited for spotting Jupiter-sized exoplanets around low-mass high-velocity stars. On the other hand, RV surveys are able to observe not only massive planets, but also Earth-sized planets orbiting massive high-velocity stars provided that 
the duration of the survey is large enough and that the Doppler signal is measured several times. Thus, a combination of Doppler spectrographs, such as \textit{G-CLEF} and \textit{ESPRESSO}, working together with \textit{TESS} will hopefully lead to the discovery of planets around high-velocity stars and consequently result in new understandings of planetary formation, evolution, and survivability.

\section*{Acknowledgments}

We thank Avi Loeb, Matt Holman, Dimitar Sasselov, Debra Fischer, and our anonymous referee for insightful comments on the paper. 
GF acknowledges hospitality from the Institute for Theory and Computation at the Harvard-Smithsonian Center for Astrophysics, 
where the early plan for this work was conceived. GF acknowledges Sapienza University of Rome for the grant 
"Progetti Avvio alla Ricerca" (Starting Research Grant, 0051276), which funded part of this work. 
IG was supported in part by funds from Harvard University and the Institute for Theory and Computation at the Harvard-Smithsonian Center
for Astrophysics.

\label{lastpage}

\end{document}